\documentclass[12pt,prd,nofootinbib]{revtex4-2}

\usepackage{graphicx}

\usepackage{amstext}
\usepackage{amssymb}
\usepackage{amsmath}

\usepackage[colorlinks=true, allcolors=blue]{hyperref}

\newcommand{\braket}[2]{\langle #1 | #2 \rangle}
\newcommand{\bra}[1]{\left\langle #1 \right|}
\newcommand{\ket}[1]{\left| #1 \right\rangle}
\newcommand{\vev}[1]{\left\langle #1 \right\rangle}

\begin{document}

\title{One dimensional Staggered Bosons, Clock models and their non-invertible symmetries.}
\author{David Berenstein$^{1,2,3}$, P. N. Thomas Lloyd$^{1}$\footnote{plloyd@ucsb.edu}}
\affiliation{$^1$Department of Physics, University of California Santa Barbara, Santa Barbara, California 93106, USA\\
$^2$ Institute of Physics, University of Amsterdam, Science Park 904,
PO Box 94485, 1090 GL Amsterdam, The Netherlands\\
$^3$ Delta Institute for Theoretical Physics, Science Park 904,
PO Box 94485, 1090 GL Amsterdam, The Netherlands
}

\date{\today}
\begin{abstract}
We study systems of staggered boson Hamiltonians in a one dimensional lattice and in particular how the translation symmetry by one unit in these systems is in reality a non-invertible symmetry closely related to T-duality. We also study the simplest systems of clock models derived from these staggered boson Hamiltonians. We show that the non-invertible symmetries of these lattice models together with the discrete ${\mathbb Z}_N$ symmetry predict that these are critical points with a $U(1)$ current algebra at $c=1$ and radius $\sqrt{2N}$ whenever $N>4$. 
\end{abstract}

\maketitle

\date{\today}

\section{Introduction}

The idea of staggered bosons was introduced in \cite{Berenstein:2023tru} as a set of bosonic degrees of freedom that are slightly delocalized, in a vein similar to staggered fermions \cite{Kogut:1974ag}. The main idea is to have a single bosonic variable per site (only one $q$, rather an $x,p$ pair of phase space) and to place the non-trivial Poisson brackets on the links of the lattice rather than directly on the sites of the lattice.

The simplest model of a staggered boson Hamiltonian in $1+1$ dimensions was shown to lead to a gapless theory in the large volume limit with central charge $c=1$. The model was shown to be robust against both noise and deformations of the quadratic Hamiltonian that include more general local terms that lead to changes in the dispersion relation \cite{Berenstein:2023tru}. The theory remains gapless. Indeed, it was argued that the staggered boson paradigm might naturally lead to gapless theories with non-trivial dynamics in a  lot of different situations. Simple examples in higher dimensions suggested a non-trivial connection with fracton models.

This stability was ascribed to the wave equation being of first order (only one time derivative) type rather than second order. In that sense, it is similar to a Dirac field.  This way  $\omega(k)$ is single valued, rather than the usual square root form that  arises from second order wave equations.
 The Fourier modes at quasimomentum $k,-k$ are conjugate to each other, so therefore $\omega(k)$, which is a continuous function of $k$, must satisfy $\omega(-k)=-\omega(k)$. These changes of sign force zero modes at $k=0,\pi$ and continuity forces $\omega(k)$ to be small in the vicinity of these points. In essence the argument described in \cite{Berenstein:2023tru} was about the simple topology of the quasimomentum circle: the first Brillouin zone of the dual lattice. In this case, $\omega(k)$ is a continuous map from the circle ($k\mod(2\pi)$) to the real numbers.

 It was also shown that translation by one unit was similar to T-duality in a related set of variables. Moreover, certain  non-linear deformations of the system that preserve the natural translation symmetry  led directly to spin chain systems at criticality  and at a self-dual coupling (the construction led to a special family of clock models that generalize the Ising model in a transverse magnetic field). In this case, the argument about the topology of the circle leading to a gapless spectrum  is not useful as the theory is not free.

 A natural question to ask is if there is a better way to describe the symmetry of the system in such a way that the non-trivial properties of the ground state and the gaplessness of the non-linear system can be explained by symmetry arguments instead. In this paper we will argue that the correct way to describe these topological properties is with the tool of non-invertible symmetries.
 In a sense, we will use the framework of the papers \cite{Cheng:2022sgb,Seiberg:2023cdc} to understand the lattice systems we have. A review of recent results on non-invertible symmetries can be found in \cite{Shao:2023gho}. This is somewhat different than the discussion based on topological lattice defects \cite{Aasen:2020jwb}, which is another way to understand non-invertible symmetries.
 We will show that this topological feature essentially arises because the staggered bosons implement some non-invertible symmetries in a very natural way: the symmetry is manifest in these variables. They arise as a symmetry of lattice translations by one unit.

 In fact,  the translation by one being similar to T-duality and constructions of the same theories based on other bosonic variables suggest that the correct translation on the lattice is by two units. This way, there is a non-trivial square root of the translation symmetry at play. Such symmetries naturally fall in the category of  non-invertible symmetries. These extend to the clock models, where this symmetry becomes the Kramers-Wannier duality of the clock models \cite{Kramers:1941kn}. The clock models with the symmetry that are constructed this way end up being exactly at the self-dual coupling, and the non-invertible symmetry is the one that implements the Kramers-Wannier duality \footnote{The Kramers-Wannier duality in the Ising model exchanges high temperature and low temperature. This is a generalization to spin chains where a coupling constant gets inverted.}.  

 The purpose of this paper is to explain in detail the paragraphs above. We will argue that the staggered boson variables make some of  these non-invertible symmetries manifest in the Hamiltonian. We then apply these ideas to the numerical studies of clock models by diagonalization at finite volume. The  non-ivertible symmetry persists and our goal is to show in these examples how one finds out what the 
 detailed physics of the critical theories are.
 
The clock models we construct depend on an integer parameter $N$ that describes the number of quantum states per site on an ordinary lattice. Their Hamiltonian is given by 
\begin{equation}
H= -\left[\sum_j g (Q_j+Q^{-1}_j) +(P_{j+1}\otimes P^{-1}_{j}+P_{j}\otimes P^{-1}_{j+1}) \right]\label{eq:clockdef}
\end{equation}
where $P_j,Q_j$ are $N\times N$ clock-shift matrices, one pair per lattice site.
The natural construction in the staggered boson variables starts with a lattice of size $2L$ and the translation symmetry on that lattice forces $g=1$.

There is a non-trivial ${\mathbb Z}_N$ symmetry that arises from the construction and our goal is to explain how this symmetry interacts with the square root of the translation symmetry $T$, which we will call ${\cal D}$ in this paper (this arises from the letter for duality).
For $N=2$, the construction gives rise to the critical Ising model in a transverse magnetic field with central charge $c=1/2$. For $N=3$, the construction leads to the tricritical Potts model with $c=4/5$. This model is solvable with parafermions \cite{Fradkin:1980th} (see also \cite{Fendley:2012vv}) and has been studied extensively in the literature. For $N=4$ the clock model we find is identical to two copies of the Ising model in a transverse magnetic field with $c=1$ \cite{Frohlich:1981yn}. For $N>4$, we argue that the model leads to a theory with $c=1$ and a $U(1)$ current algebra. The radius of the corresponding boson is caclulated in our examples to be given by $\tilde R=\sqrt{2N}$.  Our goal will be to explain why this is the correct radius by making a rather simple use of the properties of the non-invertible symmetry of the Hamiltonian \eqref{eq:clockdef}.

The paper is organized as follows. In section \ref{sec:staggered} we introduce the notion of the staggered boson in $1+1$ dimensions. We show that theses systems in a lattice of size $2L$ have
an algebra that is described by $L-1$ harmonic oscillator degrees of freedom (each count as two in phase space) plus two central elements. The translation by one is an automorphism of the algebra that exchanges the two central elements. All of the non-invertible properties of the symmetries gets tied to this non-trivial action on the central elements. We explain this in detail. Particularly, we show how translation by one is T-duality. We then explain how the clock models are built from certain gauge invariant variables under some shift symmetries of the staggered boson variables and how their original translation symmetry ends up being identified with the Krammers-Wannier duality. The natural translation invariant model  of the staggered boson variables is then found to be given exactly by the self-dual coupling of the clock model. 
Next, in section \ref{sec:numerics}, we study the clock models with ${\mathbb Z}_N$ symmetry for $N=3,4,5$. We argue how all of these lead to conformal field theories and we conjecture that when $N\geq 5$ these theories always have a $c=1$ theory with a $U(1)$ current algebra. We argue that the corresponding conformal field theory theory is a free boson at the radius $\tilde R=\sqrt{2N}$. Some of the evidence comes from the previous section which is perturbative in the original  staggered boson variables. We show that the result for the radius is exact if one understands how the non-invertible symmetries fix this self-dual point.  We also show an a-posteriori field theory argument of why the calculation doesn't apply for $N=3,4$: there are extra relevant or marginal operators that preserve the ${\mathbb Z}_N$ symmetry an the self-dual critical point. This means that certain self-dual flows away from the naive result are not forbidden. We also study deformations of the clock models at $N>4$ away from the self dual point. We argue that these give rise to a critical phase of $c=1$ conformal field theories with continuously varying critical exponents that depend on the coupling constant we vary. These phases must end in BKT transitions \cite{berezinskii1971destruction,J_M_Kosterlitz_1973}. In section \ref{sec:conc} we conclude. We also include appendices on numerical methods and finite size corrections for the finite lattice \ref{app:FS} as well as a siple proof of the equivalence of the clock model with ${\mathbb Z}_4$ symmetry at criticality  and two copies of the Ising model in a transverse magnetic field at criticality in \ref{app:twoising}.

\section{Staggered bosons}\label{sec:staggered}

Staggered bosons were introduced in \cite{Berenstein:2023tru} as a tool to write field theories in a way that blurs the usual distinctions between coordinates and their conjugate momenta.
The idea of a single staggered boson set of degrees of freedom in one dimension is as follows. We attach a single real (hermitian) bosonic coordinate $q_j$ to each lattice site in a one dimensional lattice. The commutation relations of these variables are given by
\begin{equation}
    [q_j, q_k]= i \hbar (\delta_{j,k-1}-\delta_{j,k+1})
\end{equation}
and these commutation relations are a discretized version of the continuum $U(1)$ chiral current algebra relations
\begin{equation}
[J(x), J(x')] = i \hbar\partial_x \delta(x-x') \label{eq:anomaly}
\end{equation}
The non-triviality of the construction relies on the fact that nearest neighbors talk to each other through
their Poisson brackets rather than by the shape of the Hamiltonian. 

In what follows, we will work in natural units where $\hbar=1$.
Staggered bosonic Hamiltonians are functions of the $q_j$ variables that are local (the local sum over the lattice depends on a few nearest neighbors at most). They are interesting in that they seem to lead to gapless field theories in the large volume limit very naturally. 

To build the simplest staggered boson Hamiltonian, let us use the $U(1)$ current algebra as inspiration. Consider the Sugawara form of the Hamiltonian for the continuum $J$ currents
\begin{equation}
    {\cal H}= \frac v2 \int dx J(x)^2
\end{equation}
where $v$ plays the role of the speed of light. We  can recover the equations of motion that identify $J$ as a left-moving degree of freedom from the Hamiltonian and the commutation relations \eqref{eq:anomaly}. 
The discrete version of this Hamiltonian adapted to the $q$ variables is 
\begin{equation}
 {\cal H}= \frac v2 \sum_j q_j^2\label{eq:ham1}
\end{equation}
For simplicity we will assume that we are on a periodic lattice with an even number of sites, so that $q_j=q_{j+2n}$. The equations of motion of the $q_j$ are given by
\begin{equation}
\dot q_j= v(q_{j+1}-q_{j-1}) 
\end{equation}
and we notice that the right hand side is a discretized derivative operation. In the continuum and low quasimomentum limit it leads to a chiral equation of motion.

We need to make a technical note here on the usage of momentum. There are two notions of momentum that arise in this setup. The first one, quasimomentum on the lattice, will always be referred to as quasimomentum. When taking the continuum limit we might call this spin. The reason for this is that we will implicitly use the  conventions of conformal field theory on the circle, where the notion of momentum is quantized and via the operator state correspondence becomes the spin of the corresponding operator. Momentum itself will be reserved to momentum in target space. This might be an exact or an emergent symmetry and is usually associated to certain charged vertex operators in the presence of a $U(1)$ current algebra.
This is the familiar language of string theory textbooks \cite{Polchinski:1998rq}.  Because we will find in some models such a conformal field theory, we will reserve that notation to those setups.  In other setups, the momentum quantity might instead be called charge. We also have an exact ${\mathbb Z}_{N}$ symmetry and we want to call charge the discrete charge associated to this symmetry. 

The  staggered boson set of variables is a set of generators of an  algebra. The algebra is generated by the $q_j$ variables associated to the lattice (we can take this algebra to be polynomials of $q$ for example, or some other form of convergent power series in $q$).
The algebra has two central elements when the number of lattice sites is even. These are $C_{even} = \sum_{j=0}^{n-1} q_{2j}$ and $C_{odd}=\sum_{j=0}^{n-1} q_{2j+1}$. The rest of the $2n-2$ variables give rise to a non-degenerate Poisson bracket (set of commutators) that can be described as $n-1$ copies of the Heisenberg algebra. 
If on the other hand we choose antiperiodic boundary conditions, then the staggered bosons are equivalent to $n$ copies of the Heisenberg algebra (the central elements disappear).
The fact that there are central variables (constants of motion) is the most important property we need to highlight. The fact that the algebra has a non-trivial center means that quantization (in the sense of representation theory of the algebra) is very non-trivial.

Essentially, a physical theory is not just an algebra of observables. We also need a representation of the algebra. For example, it is not enough to have angular momentum commutation relations to specify that we have a spinning particle, but we also need to say what spin a particle has.  Alternatively,  in other similar examples, we need to specify which representations of angular momentum are allowed to describe the excitations of a system. 

For the $q$ variables, there is a part of the representation that is relatively straightforward: an irreducible representation of the Heisenberg algebra is essentially unique, thanks to the Stone-Von Neumann theorem. Basically, we can describe the $n-1$ copies of the Heisenberg algebra as a set of $n-1$
harmonic oscillators. If we choose these carefully, they are the normal modes of excitations of the system.
Because the Hamiltonian \eqref{eq:ham1} is translation invariant, these normal modes are the non-trivial  Fourier modes of the $q$. For these normal modes $\tilde q_{k}$ and $\tilde q_{-k}$ are conjugate to each other.

The modes at quasimomentum $k=0,\pi$ are not included as harmonic oscillators: they would be self conjugate, but such property is not allowed for bosons.
Instead, they are represented by the two linear combinations $C_{even}\pm C_{odd}$, which are the central elements of the algebra.

On an irreducible representation $C_{even}$ and $C_{odd}$ must act as c-numbers, which are required to be real as the $C$ are hermitian (self-adjoint). Irreducible representations are therefore labeled by pairs $(C_{even},C_{odd})$. Moreover, a theory will in general be described by many of these representations, so the Hilbert space will be a direct sum of sectors where each $(C_{even},C_{odd})$ is fixed and each sector will have a set of oscillators whose frequencies are independent of the sector. The zero modes contribute to the energy an amount equal to
\begin{equation}
E_{zero}= \frac v{2n} C^2_{even}+\frac v{2n} C_{odd}^2
\end{equation}
which is obtained by looking for (classical) solutions where the zero modes are excited and nothing else is. These are given by $q_{2j}= C_{even}/n$, $q_{2j+1}= C_{odd}/n$.

The whole theory of representations of the staggered boson algebra reduce to the problem of what are the allowed values of $C_{even},C_{odd}$. The rest follows from the harmonic oscillators of the normal modes. 
The modes $C_{even},C_{odd}$ should be thought of as topological charges, and the different values of pairs of $(C_{even},C_{odd})$ should be thought of as superselection sectors determined by some additional topological considerations outside of the algebra itself.  What are the appropriate ways to do this in general is beyond the scope of the present paper. These require input beyond the algebra. 

To make contact with a standard free boson in one dimension, we will assume that the  possible values of $C_{even},C_{odd}$ form a lattice and that their quantization is independent of the size of the system.
Moreover, we will assume that for each value $(C_{even},C_{odd})$ there is a unique irreducible representation of  the algebra appearing in the Hilbert space of states.
Technically, the Hilbert space is then of the form
\begin{equation}
    {\cal H} = \bigoplus_{m,w\in {\mathbb Z}} {\cal H}_{m \vec C_1+ w\vec C_2}
\end{equation}
where $\vec C_1, \vec C_2$ are some generating vectors for a two dimensional lattice of possible values of $(C_{even},C_{odd})$ values.

The main reason to require this property is that it is possible to map a system of regular bosons in a lattice with $n$ sites, and variables $p_j, x_j$ with canonical commutation relations to the algebra of the $q$ operators. A map to the $q$ variables is given by
\begin{eqnarray}
    q_{2j}&=& p_j\\
    q_{2j+1}&=& x_{j+1}-x_j \label{eq:traditional}
\end{eqnarray}
where the subindices of the $x,p$ variables are defined only modulo $n$, rather than modulo $2n$. Under this map, we have that $C_{even}=P_{total}$ is the total momentum of the variables \footnote{This is momentum in target space, not the lattice.} and $C_{odd}\equiv 0$. We can relax this property if we allow winding $x_{n} \equiv x_0 + C_{odd}$, so that $C_{odd}$ can be thought of as the winding variable of the $x$ coordinates.

If we allow the $x$ to be globally periodically identified together as $x\equiv x + 2\pi R$ for all $x_j$ simultaneously, where $R$ is the radius of the boson, then $P_{total}$ is quantized as $m/R$ with $m$ an integer. Similarly, the winding contribution to $C_{odd}$ is quantized in integer multiples of $2\pi R$. 

The variables $q_{2j+1}$ are gauge invariant. We still allow the system to have a non-trivial identification of the $x$ variables when we go all the way around the circle. This non-trivial identification arises topologically from a difference in the {\em local trivialization of the $x$ variables} once we try to make it global. That is, the $x$ would be defined by patches and we would need to glue the patches together. This can always be trivialized in a simply connected domain.
The possible gluings are then classified by the {\em holonomy} around the circle when we have a non-dimply connected domain.
The momentum and winding are then allowed to take the possible values
\begin{eqnarray}
    C_{even} = m/R\\
    C_{odd}= w (2\pi R)
\end{eqnarray}

Basically, if we take $m,w$ arbitrary integers and we insist that for each $m,w$ there is only one irreducible representation of the algebra, we find that in the continuum limit we would be able to find a modular invariant partition function on a torus. For simplicity we will call  such models with these constraints {\em modular models}. Let us reiterate this, for the models described by the staggered boson construction this is a choice that can be made. There is no a-priori reason for this to be true given the staggered boson algebra of observable we have described so far. We choose it as a model of how it 
can work in practice.

\subsection{Translation is T-duality }

We can now ask if translation of the $q$ variables is a symmetry of the quantum theory built from the $q$ variables or not.
It is clear that the operation of exchanging $q_j \to q_{j+1}$ is an (outer) automorphism of the algebra of the $q$ variables: it preserves the commutation relations. We will call this operation $\cal D$ (from duality).
We need to know if this symmetry of the algebra can be extended to a symmetry of the Hilbert space of states. Basically, apart from the algebra, we need a Hilbert space on which it acts.

The issue at hand is that 
\begin{eqnarray}
     {\cal D}: C_{odd}\to C_{even}\\
     {\cal D}: C_{even} \to C_{odd}
\end{eqnarray}
so that the $\cal D$ operation exchanges the two central elements. Basically, it exchanges the notions of momentum and winding.  As such ${\cal D}$ is akin to T-duality. When we usually talk about T-duality, we do not consider it a symmetry of a theory. We consider it usually as an identification between two possible theories. In our conventions for $R$, it acts as 
\begin{equation}
R \leftrightarrow \frac{1}{2\pi R}
\end{equation}
so even though $\cal D$ is an automorphism of the algebra, it should not be immediately considered as a symmetry of a physical system built from it. The quantity $R$ has units related to $\hbar$, which we have implicitly set to one in these equations. The standard convention for $R$ in string theory would give different results. To restore some of the conventions  used in conformal field theory and string theory textbooks, we should use the definition
\begin{equation}
\tilde R= 2 \sqrt{\pi} R
\end{equation}
In this case, rescaled central elements behave like the momentum and winding in a vertex operator in conformal field theory
\begin{eqnarray}
    \tilde C_{even}&=&  m/\tilde R =   m / (2\sqrt \pi R) =  C_{even}/(2 \sqrt \pi)\\
    \tilde C_{odd} &=&   w  \tilde R/2 =  C_{odd}/(2 \sqrt \pi) 
\end{eqnarray}
and the T-duality transformation acts by sending $\tilde R\to 2/\tilde R$ as is more standard.

The square of the automorphism, which we will call ${\cal T}$, sends the algebra to itself keeping the central elements fixed. Because of the Stone-Von Neumann theorem, we can extend this operation to a unitary action on the Hilbert space of the oscillators without issue. Also, it keeps the superselection sectors of $(C_{even}, C_{odd})$ fixed. As such, the $\cal T$ operation can be easily promoted to a symmetry of the Hilbert space of states, and not just an automorphism of the algebra that leaves the Hamiltonian invariant. The translation by $2$ units in the $q$ variables is equivalent to translation by $1$ unit in the $x,p$ lattice that we mapped to the $q$. As such, we should think of ${\cal T}$ as the true translation operator.

We now want to think further about the fact that it would be very convenient if we can use $\cal D$ as a symmetry and not just as an identification of two different theories (a duality), so that we can write  
\begin{equation}
    {\cal T} \simeq {\cal D}^2
\end{equation}
The main ingredient we need is that there might be a subset of  $(C_{even},C_{odd})$ pairs (and not just the zero vector) that can be mapped into each other. Because we have insisted that $(C_{even}, C_{odd})$ form a lattice, this will occur if the quantizations of $C_{even}$ and $C_{odd}$ are related to each other by a rational number. 
In that case, the ${\cal D}$ operation can be extended as a symmetry for those pairs of momenta and winding that can be matched between the two different theories.

That is, we can think of ${\cal D}$ as a symmetry that commutes with the Hamiltonian in a subsector of the theory \footnote{The subsectors are defined by a combination of superselection sectors of the theory.}. For the other states where such a pairing cannot be found, we extend ${\cal D}$ by zero.
The ${\cal D}$ constructed this way is an example of a {\em non-invertible symmetry}. We define ${\cal D}$ as  the maximal possible non-invertible symmetry that is compatible with the modular models. This way, it is not T-duality, but an actual symmetry of the staggered boson theory. The correct equation is then
\begin{equation}
     {\cal D}^2 = {\cal T}\circ \Pi
\end{equation}
where $\Pi$ is a projector that commutes with both $\cal T$ and with the Hamiltonian. Because $\Pi$ is a non-trivial projector, it is non-invertible. That is, we can not define an inverse  $\Pi^{-1}$ such that $\Pi^{-1}\circ \Pi= 1$. Basically, ${\cal D}$ is not a symmetry in the sense of Wigner: either a unitary or anti-unitary operator acting on the Hilbert space of states.

We also have a boson parity operator $(-1)^B$ that acts by sending $q\to -q$ and is trivially a symmetry. 
We can also consider a space parity $\cal P$ operator as in \cite{Berenstein:2023tru} sending 
$q_j \to (-1)^j q_{-j}$ on the lattice. For most of the problems in this paper we will not delve into the details of boson parity or space parity.

\subsection{Clock models}\label{sec:clockdef}

Clock models are derived from the $q$ variables, by gauging translations of the individual $q_j \equiv q_j+\alpha$ with the same $\alpha$-periodicity for all $j$. The idea is to be able to keep the translation symmetry ${\cal T} $ and the ${\cal D}$ operation as automorphisms of the algebra that respects the  gauging.
The gauge invariant variables are then 
\begin{equation}
    K_j = \exp( 2\pi i q_j/\alpha)= \exp(i \beta q_j)
\end{equation}
and their inverses $K_j^{-1}$ where we have introduced a new parameter $\beta=2\pi/\alpha$ for simplicity. Since $\beta$ is real if $\alpha$ is real, the $K$ are unitary operators rather than hermitian  ones. Unitarity is imposed by $K_j^\dagger= K_j^{-1}$.

The idea is then to write Hamiltonians so that they are  simple algebraic expressions of the algebra generated by the $K_j$ rather than the $q$. The ${\cal D}$ and ${\cal T}$ operation work in an obvious way with these variables and produce automorphisms of the algebra of the $K$ operators.

The simplest such Hamiltonian is to take
\begin{equation}
H= \sum_j -(K_j+K_j^{-1}) \label{eq:Ham0}
\end{equation}
summed over the lattice sites.
The sign is chosen for convenience, so that in a semiclassical setting the ground state occurs when $q\simeq 0$ as a c-number and therefore $K_j=1$.

We can directly check that the $K$ variables satisfy a simple algebra
\begin{equation}
    K_j K_{j+1} = \exp(i \beta q_j)\exp(i \beta q_{j+1})= \exp( i \beta (q_j+q_{j+1}) - \beta^2 [q_j,q_{j+1}]/2) 
\end{equation}
This way their multiplication differs by a phase
\begin{equation}
    K_j K_{j+1} = K_{j+1} K_j \exp( - i \beta^2) 
\end{equation}

We will impose an additional constraint on $\alpha$. We want to set it up so that $K_j^N$ is central for some $N$ and that this $N$ is the minimal possible such value. 
This simplifies the representation theory of the Hilbert space generated by the $K$. With this constraint the $K$ can act on a Hilbert space of finite dimension, rather than an infinite dimensional one. The reason is that in an irreducible representation the quantity $K_j^N$ is  a c-number for each $j$, so polynomials of $K$ are truncated at level $N$ for each $K$. That is, the algebra generated by $K$ is a finite dimensional vector space over the complex numbers. 
This can only occur if the following {\em quantization} rule happens $  N \beta^2= 2 \pi$.
Let us convert back to the $\alpha$ parameter. We get that 
\begin{equation}
    \alpha = 2 \pi/\beta =  \sqrt{2\pi}\sqrt{N} \label{eq:radius}
\end{equation}

Let us identify for a moment this number with the possible radius of  compactification of a boson as we had before. We do this by setting $\alpha=2\pi R_N$. We would get then that 
\begin{equation}
     R_N = \sqrt{N}/\sqrt{2\pi}
\end{equation}
Alternatively, in the CFT normalization of the radius $\tilde R_N$, we would get that
\begin{equation}
    \tilde R_N=  \sqrt{ 2 N}
\end{equation}

Notice that in principle the quantization of $C_{even}$ and $C_{odd}$ arising from the gauging have the same periodicity. The spectrum would thus in principle be self-dual. However, the radius $\tilde R_N$ is not the self-dual radius of T-duality. What that means is that a model with all of those periodicities could be self-dual, but not modular invariant. We will return to this point later on. If we insist on modularity, we need to do something more sophisticated. The clock models we study do this on their own: they treat $C_{even}$ and $C_{odd}$ slightly differently.

We can also consider how this construction would work with the traditional fields as in \eqref{eq:traditional}. In that case, we can lift the gauged symmetry action on the $q_j$ variables to the $p,x$ variables that map to them, so that 
\begin{eqnarray}
p_k\equiv p_k+ r_1 \alpha \\
x_k \equiv x_k +r_2 \alpha
\end{eqnarray}
where $r_1, r_2$ are arbitrary independent integers. What this means is that the $x,p$ phase space variables belong to a 2-torus. The 2-torus is a symplectic manifold and  now has finite area. This area must be quantized by the Dirac quantization condition. This condition is equivalent to asking that the area is a multiple of $2\pi$ 
\begin{equation}
    \alpha^2 = 2\pi N
\end{equation}
basically, each semiclassical state occupies $2\pi\hbar$ quantum of area and there are $N$ such states.
Notice that this coincides exactly with the quantization \eqref{eq:radius}.

The quantization of this phase space produces a Hilbert space of dimension $N$. The two operators
$\exp(i \beta x), \exp(i\beta p)$ are gauge invariant. They must be unitary. With $\exp(i N \beta x)$ central, we can choose $\theta$ phases such that not only is the result in an irreducible proportional to the identity, but that it is the identity itself.
\begin{equation}
    \exp(i N \beta x-iN \theta)= 1
\end{equation}
We can do the same for $\exp(i \alpha p)$.
We absorb these phases into the definitions of the exponentials.
 With these, the algebra of a single site works as follows.
 \begin{equation}
     P\simeq \exp(i\beta x), Q \simeq \exp(i \beta p)
 \end{equation}
These satisfy $P^N=Q^N=1$, $P^*=P^{-1}$, $Q^*=Q^{-1}$ and more importantly
\begin{equation}
PQ = \omega QP
\end{equation}
where $\omega$ is a primitive $N-th$ root of unity $\omega= \exp(2\pi i/N)$.

The matrices $P,Q$ constructed from this algebra are called the Clock-Shift matrices of t' Hooft. They are also the matrices that describe a fuzzy 2-torus geometry.
Their algebra is such that there is a unique unitary irreducible representation of the $Q,P$ algebra. Basically, the $P,Q$ algebra are equivalent as an algebra to the set of $N\times $N matrices  $M_{N\times N}({\mathbb C})$. This algebra of $N\times N$ matrices is such that all automorphisms of the algebra are inner automorphisms (can be realized by conjugation). With the adjointness condition, these are all unitary automorphisms. 

We now have one of these per site of the lattice of the $x,p$ variables.
The new variables $K_j$ that are derived from these are given by
\begin{eqnarray}
K_{2j} &=& \exp(i \beta q_{2j}) = \exp(i \beta p_j) = Q_j\\
K_{2j+1}&=& \exp(i \beta q_{2j+1})=\exp(i \beta (x_{j+1}-x_j)) = P_{j+1}\otimes P^{-1}_{j}
\end{eqnarray}
where we have a set of $P,Q$ matrices per site. The $P,Q$ at different sites commute, because the $x,p$ at different sites do so as well.
The Hamiltonian \eqref{eq:Ham0} is then given by
\begin{equation}
H= -\left[\sum_j (Q_j+Q^{-1}_j) +(P_{j+1}\otimes P^{-1}_{j}+P_{j}\otimes P^{-1}_{j+1}) \right] \label{eq:HamPQ}
\end{equation}
These Hamiltonians are called clock models and have been studied in the literature in various papers.
This construction of the models is an independent point of view. 
Notice that similar to the $x,p$ variables, the $P,Q$ variables lead to a unique Hilbert space: the unique irreducible representation of the algebra. 

The choice of Hamiltonian \ref{eq:HamPQ} is also constrained by the boson parity $(-1)^B$, which acts as $P\to P^{-1}$ and $Q\to Q^{-1}$. Since the coefficients of $Q,Q^{-1}$ and of $P\otimes P^{-1}$ are real, we are imposing the boson parity symmetry as well. The reflection parity ${\cal P}$ acts by sending $P_j\otimes P_{j+1}^{-1}\to P_{-j}\otimes P_{-j-1}^{-1}$ and sending $Q_j \to Q_{-j}^{-1}$.

The equivalent of the zero modes consist of the product operators
\begin{eqnarray}
    Z_{even}= \prod_j K_{2j}&\simeq& \exp(i \beta C_{even})\\
    Z_{odd} = \prod_j K_{2j+1} &\simeq &\exp(i \beta C_{odd})
\end{eqnarray}
These satisfy $Z_{even}= \prod_{j} Q_j$, $Z_{odd}=1$ when written in the $Q,P$ variables and $Z_{even}^N=1$. Therefore they give rise to a ${\mathbb Z}_N$ symmetry, rather than a ${\mathbb Z}_N\times {\mathbb Z}_N$ symmetry just from the definition of $Z_{even}$ and $Z_{odd}$. Rather than an additive quantum number for a symmetry, it is a multiplicative. 

The ${\cal D}$ operation in these variables is given by
\begin{equation}
\begin{matrix} P_{j}\otimes P^{-1}_{j-1}&\ & \to&\ & Q_j\\
Q_j &\ &\to& \ & P_{j+1}\otimes P^{-1}_{j}
\end{matrix}
\end{equation}
which we recognize as the Kramers-Wannier duality. This is considered also as a non-invertible symmetry in the literature. Here we see that Kramers-Wannier duality can be thought of as being essentially T-duality in disguise.
A more general clock model is given by
\begin{equation}
H= -\left[\sum_j g (Q_j+Q^{-1}_j) +(P_{j+1}\otimes P^{-1}_{j}+P_{j}\otimes P^{-1}_{j+1}) \right]
\end{equation}
where $g$ is a real coupling constant. The value $g=1$ is considered the self-dual radius for the Kramers-Wannier duality and is supposed to correspond to a critical CFT in the infrared. It is a genralization of the Ising model in  a transverse magnetic field. Here the terms with $P\otimes P^{-1}$ are like the Ising interaction, and the terms with $Q,Q^{-1}$ are like the transverse magnetic field. In the case $N=2$, the $P,Q$ matrices become Pauli matrices and the model at $g=1$ is exactly the critical point of the Ising model in a transverse magnetic field.

In the staggered boson description the effective Hamiltonian can also be written as
\begin{equation}
H= -\sum_j 2\cos(\beta q_j)\simeq \sum(-2+\beta^2 q^2+ O(q^4) )\label{eq:naive}
\end{equation}
If $\beta$ is sufficeitnly small, we expect that the higher order terms become irrelevant in the infrared, as they are small at the $UV$ scale of the lattice and these are irrelevant operators (products of currents). How small $\beta$ has to be needs to be determined. In this sense, we expect that the perturbation theory would predict that the infrared theory is a boson with $c=1$ and some periodic radius of identification which we have suggested in equation \eqref{eq:radius} but have not determined from the full lattice model. Part of our goal is to understand this: for what $N$ does the more naive staggered boson truncated at the quadratic order in \eqref{eq:naive} gives already a good approximation to the physics? Basically, large modifications in the UV might take us away from the basin of attracting on an infrared fixed point.

In the rest of the paper, we will explore the clock models constructed this way at the critical coupling by studying them numerically and also by making sense of the non-invertible symmetry to explain their properties.

\section{Numerical results for clock models at the self-dual point for $N=3,4,5$.}\label{sec:numerics}

Let us begin studying the various clock models numerically. The idea is to do an exact diagonalization of the Hamiltonian at finite volume and to extrapolate to the large volume limit. We have $L$ sites of $P,Q$ variables. 
Translation invariance is an exact symmetry with generator $T$, and $T^L=1$. The possible eigenvalues are 
$\exp( 2\pi i k/L)$. The quantity $k$ is an integer. We call the quantity $2\pi k/L$ the quasimomentum. It is a periodic variable with period $2\pi$. We also have an exact ${\mathbb Z}_N$ symmetry that commutes with $T$.

The number of states per lattice site is $N$. Therefore the Hilbert space is of size $N^L$. Before we diagonalize, we split the Hilbert space into sectors of fixed quasimomentum and fixed charge. This reduces the problem to diagonalizing matrices of sizes that are roughly  $N^L/(N L) \times N^L/(N L)$, with $k$ fixed and fixed charge. To fix the charge of the states, we choose a basis where $Q$ is diagonal, that is, we choose
\begin{equation}
    Q_j\equiv \hbox{diag}( \exp(2\pi i q_j/N))
\end{equation}
where $q_j$ in an integer modulo $N$. The basis that diagonalizes $Q$ will be obtained by using the label $q_j$ at each site. This way, the list of states is a collection of integers $(q_0, \dots, q_{L-1}) $ and the total charge is $Q=\exp( 2\pi i \sum_j q_j/L)$, or if we use additive notation 
\begin{equation}
q=\sum q_j \mod(N).
\end{equation}
For simplicity, we use $q\in {0, \dots L-1}$.
To diagonalize $T$, we order the states lexicographically. We form equivalence classes of states by translation. For example the two states $\ket{1010}=\hat{T}\ket{0101}$ would be equivalent.
We thus define a representative state $\ket{n^\prime}$ which is the first one appearing in the lexicographic order, which represents all the translationally equivalent states $\{\ket{n},\hat{T}\ket{n},...,\hat{T}^{L-1}\ket{n}\}$.  We also define the period of $\ket{n^\prime}$, $p(n^\prime)$, as the number of applications of $\hat{T}$ necessary to recover the initial state. That is,  $\hat{T}^{p(n^\prime)}\ket{n^\prime}=\ket{n^\prime}$. This must necessarily be a divisor of $L$. The analysis therefore simplifies if $L$ is a prime number, as there are very few states with period $1$.

Once this subset of states are identified,  we construct the eigenvectors of the different quasimomentum blocks as follows
\begin{equation}
    \ket{k,q}_{n^\prime}=\frac{\sqrt{p(n^\prime)}}{L}\sum_{j=0}^{N-1}\exp(-2\pi i k/L) \hat{T}^j\ket{\theta}_{n^\prime}.
\end{equation}
The $\sqrt{p(n')}$ ensures that these states are orthonormal. That is, the overlap of the states satisfies $\braket{k,q,n'}{\tilde k, \tilde q ,\tilde n'}= \delta_{k,k'}\delta_{q,q'}\delta_{n,n'}$. For convenience, we also define the roots of unity $\omega_k= \exp(2\pi i k/L)=\omega_1^k$.
We construct the entry of each block $\bra{k,q}_{m^\prime}\hat{H}\ket{k,q}_{n^\prime}$ where $m^\prime$ and $n^\prime$ simply indicate different states within the basis shown above as follows:
\begin{eqnarray*}
\bra{k,q}_{m^\prime}\hat{H}\ket{k,q}_{n^\prime}&=&
\bra{k,q}_{m^\prime} \frac{\sqrt{p(n^\prime)}}{L}\sum_{j=0}^{L-1}\omega_k^{-l}\hat{T}^j\hat{H}\ket{q}_{n^\prime}\\
&=&\bra{k,q}_{m^\prime} \frac{\sqrt{p(n^\prime)}}{L}\sum_m\sum_{j=0}^{L-1}\omega_k^{-l}\hat{T}^jH_{mn^\prime}\ket{q}_m\\
&=&
\bra{k,q}_{m^\prime} \frac{\sqrt{p(n^\prime)}}{L}\sum_m\sum_{j=0}^{L-1}\omega_k^{-l}\hat{T}^jH_{mn^\prime}T^{d(m)}\ket{q}_{m^\prime}\\ &=&
\bra{k,q}_{m^\prime} \frac{\sqrt{p(n^\prime)}}{L}\sum_mT^{d(m)}H_{mn^\prime}\sum_{j=0}^{L-1}\omega_k^{-l}\hat{T}^j\ket{q}_{m^\prime}\\ &=&
\bra{k,q}_{m^\prime} \sqrt{\frac{p(n^\prime)}{p(m^\prime)}}\sum_m\omega_k^{d(m)}H_{mn^\prime}\ket{k,q}_{m^\prime}\\
&=&\sqrt{\frac{p(n^\prime)}{p(m^\prime)}}\sum_m^*\omega_k^{d(m)}H_{mn^\prime}
\end{eqnarray*}
where $T^{d(m)}\ket{m^\prime}=\ket{m}$ and the sum over $m$ to $*$ is the sum over all states translationally equivalent to $\ket{m^\prime}$. The matrix elements $H_{mn^\prime}=\bra{m}\hat{H}\ket{n^\prime}$ have $\bra m$ in the original natural basis of the Hilbert space, rather than the one that we built by translations.
Given the matrix elements $\bra{k,q}_{m^\prime}\hat{H}\ket{k,q}_{n^\prime}$ are in block form, 
 we then numerically calculate the eigenvectors and eigenvalues for each block thus providing the spectrum of the Hamiltonian.

The large volume limit is obtained by taking $L\to \infty$ keeping $k$ fixed and zooming in to the eigenvalues of the energy near the ground state. We rescale this limit to a circle of radius $r=1$. The integer $k$ becomes the continuum momentum on the circle in units of the inverse radius $1/r$. We call this quantity spin as is common in conformal field theory. We are reserving the notion of momentum for target space.  The limit also requires us to rescale the energy operator relative to the vacuum, so that $H_{eff}= A\Delta H+B$, where $A,B$ needs to be adjusted. The values $\Delta H$ in a critical (conformal) point are expected to be of order $1/L$. We want the values of $H_{eff}$ to be of order $1$. Therefore $A$ needs to be of order $L$ and $B$ is of order one. The quantity $B$ controls the zero point energy. The quantity $A$ controls the energy of the excitations (the gap).
The question is then how to normalize $H_{eff}$ correctly. 

We will do this by the Koo-Saleur procedure \cite{Koo:1993wz}. The idea is that in the large volume limit, the Hamiltonian will become a continuous integral of a local Hamiltonian
\begin{equation}
H_{eff}= \int d\theta {\cal H}(\theta)
\end{equation}
which must be the Hamiltonian of a conformal field theory with $H= L_0+\bar L_0+{\cal E}_0$ where $\cal E_0$ is the zero point energy of the conformal field theory. In the large volume limit we have a conformal field theory with a symmetry controlled by two copies of the Virasoro algebra: one for left movers and another for right movers. The Virasoro algebra is an emergent symmetry. In the lattice we have discrete space translation and continuous time evolution. There is no continuous momentum on the lattice: it is discrete at finite volume.
The only generator that survives as a continuous symmetry this way is the Hamiltonian. 

In the large $L$ limit discussed above, we should also have operators that represent the generators of the Virasoro algebra. 
Here, the correct normalization (units) of $H$ will be set by the Virasoro algebra:
\begin{equation}
    [L_n,L_m]= (m-n) L_{m+n}+ \frac c{12}  n(n-1)(n+1) \delta_{m+n,0}
\end{equation}
The quantity $B$ is tuned so that $H_{eff}$ gives the correct zero point energy and the $A$ is normalized from commutators of the Hamiltonian with the $L_{n}$ (in practice, this can be thought of as a Ward identity for the symmetry). In particular, if we have the $L_n$, the energy spectrum will be integer spaced representation by representation. If we have a preferred set of states where we know $\Delta H$ and $\Delta H_{eff}$, we can numerically find the normalization of $A$. The quantity $c$ is the central charge of the conformal field theory. This is a property of the infrared field theory that needs to be deduced from the numerical computations.

The idea is that we need some way to find some information about the $L_{n}, \bar L_n$ from the lattice. To do so, we want to identify the local Hamiltonian  with ${\cal H}\equiv T_{++}+T_{--}= T+\bar T$, the $T_{tt}$ component of the stress tensor of the conformal field theory on the circle. The Fourier modes of ${\cal H} $  are thus a linear combination of the Fourier modes of $T_{++}, T_{--}$, which generate the two copies of the Virasoro
algebra. These give
\begin{equation}
    H_{n}= \int d\theta \exp(i n\theta) {\cal H}(\theta) = L_n+\bar L_{-n} \label{eq:vir0}
\end{equation}
The two copies of the Virasoro modes are $L_{n}, \bar L_n$. They appear with opposite signs in \eqref{eq:vir0} as the $\pm n$ Fourier modes are correlated with positive or negative frequency depending on if the corresponding mode is left moving or right moving.

We approximate these linear combinations by the discrete Fourier transforms of the local Hamiltonian on the lattice. We also take the spin $L_0-\bar L_0$ to be given by the quasimomentum $k$ ($k$ fixed occurs for both positive and negative $k$). The idea is then that the Virasoro algebra commutators will be realized up to finite size (scaling) effects on the lattice and partial information of the Virasoro generators can be obtained from the $H_n$ themselves. Only in the limit $L\to \infty$ one should recover the exact Virasoro algebra. 

The Conformal Field Theory has a unique vacuum state $\ket 0$ that is annihilated by the $L_n, \bar L_{n}$ with $n\geq 0$.
It is also annihilated by $L_{-1}, \tilde L_{-1}$ (it is invariant under $SL(2)$). The Hamiltonian includes a zero point energy equal to ${\cal E}_0=-c/12$ (the Cassimir energy on the circle), but that contribution does not belong to $L_0,\bar L_0$.
The descendants under the Virasoro symmetry will have actions of $L_{-n}$ acting on various states. 
These can be obatined by acting directly with the modes $H_n$, as the vacuum (or any primary) is also annihilated by $L_{n}$ with $n>0$. 
For example
\begin{equation}
H_n\ket 0 \simeq \left\{\begin{matrix}\bar L_{-n}\ket0&\hbox{If $n>0$}\\
L_{n}\ket0&\hbox{If $n<0$}
\end{matrix}\right.
\end{equation}
Acting with $L_{-n}$ increases the energy and the spin by $n$, and acting with $\bar L_{-n}$ increases the energy and reduces the spin by $n$.

The first non-trivial descendant of the vacuum occurs at a shift in energy of $2$ and at spin $\pm 2$.
Under the operator state correspondence these would be the two operators $T,\bar T$ determining the continuum stress tensor currents. 
The idea is then to normalize as follows: first, find the ground state.
Check that it is in the $k=0$ sector.

One can then find the lowest energy state with spin two. That is, look at state at $k=\pm 2$. Identify these excited states with the $T,\bar T$ states at energy two and spin $2$. Use the $\Delta H$ from the numerical calculation and set the units for $A$ so that the energy difference is numerically identical to two in $H_{eff}$. This is simple and based entirely on the vacuum. However, it not ideal as the finite size corrections might be too large.
A better way to fix the splitting is to take the lowest non-trivial primary and its first descendant.
Their difference in energies is one (as opposed to two) and the spin difference is one, rather than two. It has smaller finite size corrections as well. 
In the appendix \ref{app:FS} we describe how to deal with finite size corrections more carefully to obtain a better determination of the correct physics of the conformal theory and what is the procedure we follow in more detail.

The Fourier modes for $H$ are defined by 
\begin{equation}
\begin{aligned}
\tilde{H}_{n} =-\frac{N}{2 \pi} \sum_{j=0}^{L-1}\left[e^{i\left(j+\frac{1}{2}\right) n \frac{2 \pi}{L}}\left(P_{j} P_{j+1}^{\dagger}+\text { h.c. }\right)\right.\\ \left.+e^{i j n \frac{2 \pi}{L}}\left(Q_{j}+\text { h.c. }\right)\right]
\end{aligned}
\end{equation}
notice that we have improved the operator by taking into account that the staggered boson construction has a non-trivial square root of the translation operator, so some elements have an extra shift of a half unit of the lattice spacing. Their action on states can be readily computed. 

To understand how to compute the central charge $c$, once we have normalized the Hamiltonian units correctly we utilize the Virasoro algebra as follows:
\begin{equation}
\bra{0}H_{2}^\dagger H_{-2}\ket{0}=\bra{0}L_{2}L_{-2}\ket{0}=\bra{0}[L_{2},L_{-2}]\ket{0}=||H_{2}\ket{\mathbf{1}}||^2=\frac{c}{2}.
\end{equation}
The theory of descendants for the vacuum $\ket 0$ on the lattice will proceed by acting with the $\tilde H$ operators instead of the Virasoro operators of the continuum. A different determination of $c$ can be done with the Cassimir energy (this is also explained in the appendix \ref{app:FS}). 

We can now organize all our states by their energy, and their spin. The charge quantum numbers are also important but they do not pertain to the Virasoro algebra. In the continuum limit, the spectrum will be organized by unitary irreducible representations of the Virasoro algebra. When the central charge $c<1$, these are Verma modules (they might have some null states that need to be removed). States will be primary if they are annihilated by all the $L_{n}, \bar L_n$ with $n>0$. An easier way to identify them is that they are the lowest energy states in their Verma module. All other states are descendants and are produced by acting with products of $L_{-k}$. These actions increase the energy and spin by $k$ units. 
For low lying states, only the first few $L_{-k}$ matter. 
When we are on a lattice, these descendants are only an approximate notion. We build them by acting with the lattice Fourier modes and  then projecting them to the specific eigenstates of the Hamiltonian that are closer to them in norm (the ones with large overlap).
Additional primaries are identified by being eigenstates of the Hamiltonian that are orthogonal to the states that are generated by the lattice $L_{-k},\bar L_{-k}$ after they have been projected on eigenstates of $H$ and that are also lower in energy than the other descendants that have been produced so far. This way, one gets an approximate table of primaries (with their spin and energies) and an approximate value of the central charge. 

There are then two simple ways to define the conformal dimension of the operators. If we have the energy and the spin $\Delta,S$, then the conformal weights can be computed by solving $\Delta= \Delta H_{eff}= h+\bar h$
and the spin $S=h-\bar h$.
Alternatively,  we can use any of a number of definitions that arise from Virasoro generators acting on primaries and using the commutation relations
\begin{eqnarray}
h_1 &=&\frac 12 (||L_{-1}\ket{\psi}||^2)\\
h_2&=&\frac{1}{4}\left(||L_{-2}\ket{\psi}||^2-\frac{c}{2}\right)
\end{eqnarray}
either of which can be used as a definition of $h$. Since these arise from the exact commutation relations of Virasoro generators, one is testing the Ward identities of the Virasoro algebra. We will use these conventions in the tables. 

In one of the definitions we are measuring directly the energy and the spin, and in the other case we are using Ward identities of the symmetry to determine the quantities. In the large L limit both give the same answer. At finite $L$ they differ by finite size corrections (these are organized in inverse powers of $L$) and the differences in energies between the states also stop being integer spaced for the same reasons. A similar computation can be done with $\bar h$ as derived from $\bar L$. 
The difference between the two ways of computing the quantum numbers gives us a handle on the expected size of the finite size corrections directly from the numerical evaluation of energies.

With these tools on hand, we proceed to study the various examples.

The clock model at $N=2$ is the Ising model in a transverse magnetic field the $c=1/2$. It is very well understood in the literature and has been analyzed numerically in many other works. It can be solved exactly with free fermions.
We will instead start our numerical studies at $N=3$.

\subsubsection{The tricritical Potts model}

The model is defined by the following $P,Q$ matrices at each site
\begin{equation}
    Q\simeq \begin{pmatrix}
        1&0&0\\
        0&\omega&0\\
        0&0&\omega^2
    \end{pmatrix}, \quad P\simeq \begin{pmatrix}
        0&0&1\\
        1&0&0\\
        0&1&0
    \end{pmatrix}
\end{equation}
and the Hamiltonian
\begin{equation}
H= -\left[\sum_j (Q_j+Q^{-1}_j) +(P_{j+1}\otimes P^{-1}_{j}+P_{j}\otimes P^{-1}_{j+1}) \right] 
\end{equation}
We will use the convention that we have $L$ sites for the $P,Q$ matrices. There is a ${\mathbb Z}_3 $ symmetry generated by 
\begin{equation}
    Z= \prod_{j=1}^L Q_j
\end{equation}
We also have a translation invariance with generator $T$ and  $T^L=1$. 

The critical model with $N=3$ is integrable. It can be solved exactly with parafermions \cite{Fradkin:1980th} (see also \cite{Fendley:2012vv}), although the prafermions are not completely well defined at finite volume with periodic boundary conditions. The model has been very well studied and one can also explicitly find a map of the local lattice operators to the conformal primaries \cite{Mong:2014ova}. 

 Instead, we follow the numerical approach we have delineated and which also follows from the Koo-Saleur analysis \cite{Koo:1993wz}. We get the following numerical results from the 60 lightest states with $E\leq 3.1$.

\begin{table}
\begin{tabular}{|l|l|l|l|l|l|l|l|l|}
\hline
{Energy level \#}& $h_{1}$ & $h_2$ &$\bar h_{1}$ &$\bar h_2$ & $\Delta=h_1+\bar h_1$ & $\Delta=\Delta E$ &Spin&Charge\\
\hline 1 &0&0 & 0&0&0&0&0&0\\
2,3 & 0.0672 &0.0672&0.0672&0.0672&0.134&0.139&0 &$\pm 1$\\
4 &0.406 &0.408 & 0.406 & 0.408&0.812 &0.859&0&0\\
9,10&.614&.643&.614&.643&1.23&1.35&0&$\pm$1\\
11,12 &.372&.369&1.285&1.287&1.657&1.76&$\pm 1$& 0\\ 43&1.324&1.319&1.324&1.319&2.648&2.868&0&0\\
49,50 & .001& .001& 2.548& 2.55& 2.549& 2.977& $\pm 3$&0\\
\hline
\end{tabular}
\caption{For $N=3$ we determine the central charge to be $c = .803 \sim 4/5$. When we have a degeneracy due to spin, we quote the numbers for the positive spin. }\label{tab:Z3}
\end{table}

The numerical data appears in table \ref{tab:Z3}. We should notice that there are two determination of the  energies (dimensions) that are recorded. One is in terms of the Ward identity, and the other one is in terms of the direct measured value of the energies. Since we expect the representation of the Virasoro algebra to be unitary, the dimension should be greater than the absolute value of the spin. Also, because quasimomentum is strictly quantized in the lattice, the notion of spin needs to be an integer. 
For example we see that there is a state with spin three and dimension almost equal to three. Unitarity requires that the dimension be greater than or equal to three. More precisely $h,\bar h$ must be greater than or equal to zero. This automatically implies that if $S=h-\bar h =3$, then $h\geq 3+\bar h$. 
If we need to identify a conformal field theory with a (chiral) primary field of dimension three and a ${\mathbb Z}_3$ symmetry, at central charge $c=4/5$, we end up with the unique choice  of the tricritical Potts model \cite{Fateev:1987vh}.  

The result of the Ward identity is smaller than the one determined from the energy. The Ward identity discrepancy is from 4\% to 10\% increasing with the energy. We take these to be a determination of the systematic error bars of the computations. That is, heavier states are less well determined than the lighter states. This is expected from the 
finite size effects discussed in the appendix \ref{app:FS}.

We now need to identify the table of primaries we found with the table of possible primaries of a $c=4/5$ conformal field theory. Since we have primaries of $(h_1,\bar h_1)=(0,3)$ (non-zero spin), we have a non-diagonal partition function. We see that the spin $3$ primary is allowed in the table of primaries. 
This means that the symmetry of the system should become the $W_3$-algebra.
We also see that the states with $h\sim 0.0672$ must be identified with representations of $h=1/15$. In this case the result of the Ward identity computation is better than the one for the energy. Similarly $h\sim 0.406$ should be identified with $h=2/5$ in the minimal model table.
The states with $h\sim 0.61\to 0.675$ should be identified with primaries of conformal weight $(2/3,2/3)$.
The first state with spin at $S=1$ has a dimension of $1.76\sim 1+ 2* 2/5$, so it has a primary of conformal weight $2/5$ and a primary of conformal weight $7/5$ glued together. The state with spin zero and energy 2.868 should be identified with 
the $(7/5,7/5)$ primary.

\subsubsection{The $N=4$ model.}

We want to do the same type of anaylsis for the $N=4$ model. As can be read from the appendix \ref{app:twoising}, the $N=4$ model at criticality is identical to two copies of the critical Ising model in a transverse magnetic field. This is a well known fact. 
In that sense, if we know the solution of the first, we know the solution of the second.
Our goal in this case is to see that the procedure outlined before for $N=3$ can also be carried out numerically on a small volume and that the results can be trusted to  find properties of the critical point directly. Thus, from the purely theoretical point of view we know that $c=1$. Because of our description of the clock model starting from the staggered boson, we expect that $c\leq 1$ and as $N$ gets larger that we are closer to a perturbative regime of the staggered boson. Here we see that since we are already at $c=1$, 
we should have that $c=1$ for $N\geq 4$.

\begin{table}
\begin{tabular}{|l|l|l|l|l|l|l|l|l|}
\hline
{Energy level \#}& $h_1$ & $h_2$ & $\bar{h}_1$ & $\bar{h}_2$ & $\Delta=h_1+\bar{h}_1$ & $\Delta E$ & Spin & Charge\\
\hline
1 & 0 & 0 & 0 & 0 & 0 & 0 & 0 & 0\\
2,3 & 0.0624 & 0.0624 & 0.0624 & 0.0624 & 0.1248 & 0.131 & 0 & $\pm 1$\\
4 & 0.125 & 0.125 & 0.125 & 0.125 & 0.25 & 0.252 & 0 & 2 \\
5,6 & 0.48 & 0.48 & 0.48 & 0.48 & 0.96 & 0.999 & 0 & $0,2$ \\
11,12 & 0.543 & 0.543 & 0.543 & 0.543 & 1.086 & 1.125 &0 &  $\pm 1$ \\
13,14 & 0.125 & 0.125 & 1.012 & 1.012 & 1.137 & 1.232 & $\pm 1$ & 0 \\
23,24 & 0 & 0 & 1.64 & 1.64 & 1.64 & 2 & $\pm 2$ & 0 \\
29 & 0.96 & 0.96 & 0.96 & 0.96 & 1.92 & 2.08 & 0 & 0 \\
50 & 1.01 & 1.11 & 1.01 & 1.11 & 2.02 & 2.3 & 0 & 2 \\
\hline
\end{tabular}
\caption{For $q=4$ we determine numerically that the central charge should be $c = .998 \sim 1$. }\label{tab:Z4}
\end{table}

From the table \ref{tab:Z4} we see that there is a primary state at energy $2$
 and spin 2 that is degenerate with the stress tensor. This is an indication that the model knows about the splitting into two separate conformal field theories at the critical point. 
 Also, the theory does not have a $U(1)$ current field. In the classification of $c=1$ conformal field theories, one has three possibilities \cite{Dijkgraaf:1987vp}: a $U(1)$ theory at radius $R$ for some $R$, an orbifold at radius $R$ (this is a ${\mathbb Z}_2$ acting on the circle target space by $\theta\leftrightarrow -\theta$ or some exceptional cases. The orbifold has two fixed points. This predicts that the ground state of the twisted sector has dimensions $(1/2,1/2)$ and is doubly degenerate (one for each fixed point of the circle). These are easily seen in the states $5,6$. The double degeneracy is counting the two fixed points of the orbifold. Notice that the degeneracy has associated to it different values of the ${\mathbb Z}_4$ symmetry. In terms of the free fermion CFT, these are the $\epsilon_1=\bar \psi_1\psi_1$ and $\epsilon_2=\bar\psi_2 \psi_2$ states (the so called energy operators). 
 Similarly, there should be operators with weight $(1/16,1/16)$ associated to the twist operators for the fermions. These are the $\sigma_1, \sigma_2$ operators of the two copies of the  $c=1/2$ field theory.
 Here we see that they have charge $\pm 1$. The state at energy $0.25$ is the product $\sigma_1\sigma_2$.

 We also see that there is no chiral state with $S=1, \Delta=1$, which is what we would expect if we had a current algebra. The numerical values are all consistent with the identification to the correct conformal field theory. Also, the degeneracies are such that in the large volume limit one should be able to recover a modular invariant partition function.

 The notion of the radius for two copies of Ising  is at $\tilde R=2$ or $\tilde R=1$. Our naive prediction based on the free field limit  was at $\tilde R= \sqrt{2*4}$, but that also assumed that the $U(1)$ current was present. Basically, at $N=4$ we're now recovering the correct central charge, but not some of the other expected features. 

\subsubsection{The $N=5$ model.}

Finally, we can analyze the $N=5$ case. The numerical results are determined from a very small lattice of size 7. Thus errors are expected to be large. Again, our prediction is that now $c=1$, which is verified by our numerical estimate of $c$. 
The dynamics can also be written in terms of parafermion variables \cite{Fendley:2012vv}.
In general, parafermion CFT's in the sense of Fateev and Zamolodchikov have central charge $c=2(n-1)/(n+2)$ where $n$ is an integer \cite{zamolodchikov1985nonlocal}. When $n=3$ it coincides with the ${\mathbb Z}_3$ model discussed above. For $n=4$ it gives rise to a $c=1$ CFT and for $n>4$ it leads to CFT's with central charge $c>1$. This means that the model can not be described as a simple parafermion theory. After all, we expect that $c\leq 1$ from our description of the system. There are integrable flows from the parafermion theory to a gapped phase \cite{Fateev:1991bv}. The flow is self-dual with respect to the Kramers-Wannier duality. One can also understand some of these with the help of statistical lattice models that can be solved exactly \cite{Fateev:1982wi}.
It is expected that the integrable flow at negative coupling leads to a non-trivial CFT. The continuum model might be integrable, but this is not expected to be so in the lattice model. We believe that this lattice model is the endpoint of the flow \footnote{We thank P. Fendley for this suggestion}.  Our analysis does not depend on that observation.

Our first task is to see if the theory we have constructed has a $U(1)$ current, by looking at the data in \ref{tab:Z5} .
If that is the case, there should be states with $E=1,S=1$. We find a pair of states that seem to to fit the requirement (states $8,9$). Again, unitarity requires that $\bar h\geq S$, so the data of $\bar h$ is low, and the one from $\Delta E$ passes the test. 

\begin{table}
\renewcommand{\arraystretch}{1.5}
\renewcommand{\tabcolsep}{4pt}
\begin{tabular}{|l|l|l|l|l|l|l|l|l|l|}
\hline
{Energy level \#} & $h_1$ & $h_2$ & $\bar{h}_1$ & $\bar{h}_2$ & $\Delta=h+\bar{h}$ & $\Delta E$ & Spin & Charge\\
\hline
1 & 0 & 0 & 0 & 0 & 0 & 0 & 0 & 0\\
2,3 & 0.05 & 0.05 & 0.05 & 0.05 & 0.10& 0.108 & 0 & $\pm 1$\\
4,5 & 0.171 & 0.17 & 0.171 & 0.17 & 0.342& 0.368 & 0 & $\pm 2$ \\
$6, 7$ & 0.442 & 0.441 & 0.442 & 0.441 & 0.884 & 0.982 & 0 & $\pm 2$ \\
$8, 9$ & 0.031 & 0.033 & 0.884 & 0.879 & 0.915 & 1.09 & $\pm 1$ & 0 \\
18,19 & 0.622 & 0.618 & 0.622 & 0.618 & 1.24 & 1.38 & 0 & $\pm 1$ \\
20 & 0.633 & 0.621 & 0.633 & 0.621 & 1.26 & 1.409 & 0 & 0 \\
45 & 0.973 & 0.949 & .973 & .949 & 1.946 & 2.15 & 0 & 0 \\
92,94 & 1.524 & 1.579 & 0.811 & 0.871 & 2.335 & 2.816 & $\pm 1$ & -2 \\
93,95 & 1.524 & 1.579 & 0.811 & 0.871 & 2.335 & 2.816 & $\pm 1$ & 2 \\
96,97 & 1.63 & 1.577 & 0.869 & 0.821 & 2.5 & 2.834 & $\pm 1$ & 0 \\
$102,103$ & 1.215 & 1.201 & 1.215 & 1.201 & 2.430 & 2.901 & 0 & $\pm 2$ \\
\hline
\end{tabular}
\caption{For $q=5$ we determine the central charge to be $c=1.01\sim 1$.}\label{tab:Z5}
\end{table}

We now want to determine what the radius of the CFT that we get, assuming that we have a current algebra. The lowest lying states that are not the vacuum and the current should be associated to vertex operators of the form
\begin{equation}
V= :\exp( i  k X):=:\exp(i k X_L) \exp(i k X_R): 
\end{equation}
where we have split the left and right moving modes.
These have conformal weights $(k^2/2,k^2/2)$. Matching what we have suggests that
$k^2/2 \simeq 0.05$. If we associate these to a vertex operator with momentum, we find that $k\sim 1/\tilde R$ and therefore
$(1/\tilde R)^2 \simeq 0.1$. Inverting, we seem to get $R\simeq \sqrt {10}$, which was our naive prediction from the staggered boson computation where we constructed the model from.
These should be double degenerate, as both $k$ and $-k$ are allowed. This is verified in the table.

If the theory is equivalent to a $U(1)$ current algebra, then there should be states at twice the momentum with four times the energy. These should be identified with states $4,5$. Similarly, if we triple the momentum, we should have states with $9$ times the energy, which we see as states $8,9$. Continuing this way we would find states with $16$ times the ground state energy (states 18,19), and $25$ times the simplest state. However at this stage since the lattice has very small volume, the quality of the states and energies is deteriorating.
Notice that the $\mathbb{Z}_5$ charge seems to be given by $k/(min(k)) \mod(5)$, that is, it is correlated with the (target space) momentum. 
We will explain this in the net section.

There are similar contributions that arise from winding modes, where $k_L=-k_R$. In a modular invariant theory at $\tilde R=\sqrt{10}$ the values of $k_L$ for winding modes would be quantized starting at $5k_{min}$, where we have taken the dual radius $2/\tilde R$ as a starting point. This is also at 25 times the energy of the simplest charged vertex operator. The state 20 in the table is suggestive, but these would also be degenerate with states that arise only from momentum.

The point is that if the radius is $R=\sqrt {10}$ there should be a four-fold degeneracy of these excited states, rather than the more standard two-fold degeneracy. States with both momentum and winding would have $k_L\simeq 6 k_{min}$ and $k_R\simeq -4 k_{min}$ for example. That is, they would have conformal weights $(36/20,16/20)$ and spin one and these seem to be given by $92,94,93,95$.

There is also the marginal operator $\partial X \bar\partial X$ which should have charge $0$ and weight $(1,1)$. The state labeled 45 seems to fit  that description.

To understand the analysis, our numerical evidence points to the conclusion that the $N=5$ theory should indeed be identified with a $U(1)$ current algebra at radius of the boson  $\tilde R=\sqrt {10}$.
The list of states that we get also seems to be compatible with modular invariance. This gives us a puzzle to solve. Our next goal will be to argue that this should be the case for all the other $N>4$ models, namely, that we get a conformal field theory at radius $\tilde R= \sqrt {2N}$ exactly. This has also been seen numerically in \cite{Li:2014pta,Li:2019dkb}, where the numerical calculations were done with a mixture of DMRG and exact diagonalization methods.

 Our goal in the next section is to explain how to arrive at this conclusion with a  non-perturbative argument.

\subsection{Non-invertible symmetries and the radius of the $U(1)$ theory.} 

Our goal is now to explain how we should think about the naive prediction of the radius that we did in section \ref{sec:clockdef}. The results of the $N=5$ case seem to give rise to the same number, but it leads to a modular invariant partition function. Our naive analysis in defining the clock models did not do that: the clock models where constructed from the $(x,p)\to q$ map, rather than just from the $q$ variables directly.

The point is that the map from $(x,p)\to q$ misses a zero mode of winding. In the corresponding $P,Q$ variables, we only have a ${\mathbb Z}_N$ symmetry generated by $Z= \prod Q_i$. The corresponding $\tilde Z= \prod P\otimes P^{-1}\equiv 1$ is identically equal to one. 
The duality operation is such that 
\begin{eqnarray}
{\cal D}\times Q \to P\otimes P^{-1} \times  {\cal D}\nonumber\\
{\cal D} \times P\otimes P^{-1}\to Q\times {\cal D}
\end{eqnarray}
and therefore can only map states with charge zero non-trivially. The projector that appears in \begin{equation}
    {\cal D}^2 \equiv T\circ \Pi
\end{equation}
should be the projector to the ${\mathbb Z}_N$ invariant states. What this means in practice is that the quantization of the zero modes of momentum $C_{even}$ and winding  $C_{odd}$ is different. 
We have a lattice $\Lambda$ of $\vec C$ values, and a sublattice $\tilde \Lambda$ of values where we can swap the $C_{even}, C_{odd}$.
The quotient  $\Lambda/\tilde \Lambda$ will give us equivalence classes by elements of $\tilde \Lambda$.
This is identical to ${\mathbb Z}_N$ which classifies the states that are not in the $\tilde \Lambda$ lattice. We see this way that ${\mathbb Z}_N$ should be identified with the momentum modulo $N p_{min}$ where $p_{min}$ is the minimum momentum.
After all, $N p_{min}$ belongs to $\tilde \Lambda$ and the states with winding $w=1$ also belong to $\tilde \Lambda$, as they are obtained by acting with ${\cal D}$ on states with momentum $N p_{\min}$.

We now use the fact that the ${\cal D}$ operation should also be understood as a T-duality. The T-duality at radius $\tilde R$ will map to a radius $2/\tilde R$. We are also quotienting by ${\mathbb Z}_N$ when we think of ${\cal D}$ as a non-invertible symmetry that must get rid of the charged states. 
When we do that, we should have that the  theory of the T-dual radius $2/\tilde R$ should be equivalent to the theory that we made by taking the quotient with respect to ${\mathbb Z}_N$. Since this keeps only the momenta that are in $\tilde \Lambda$, it is equivalent to taking $\tilde R\to \tilde R /N$. Basically, the circle should be $N$ times smaller.
That is, if the theory is modular invariant, the theory should be self-dual to an orbifold of itself.
If we equate the radii of these two ways of thinking about the non-invertible symmetry as T-duality, we find that 
\begin{equation}
    \frac{\tilde R} N= \frac  2{\tilde R} 
\end{equation}
We conclude that $\tilde R = \sqrt{2N}$ exaclty. 
This works so long as we have a $U(1)$ current algebra so that the radius and the symmetry are easily related to each other. Our conclusion is therefore that the theories with the ${\mathbb Z}_N$ symmetry are at radius $\tilde R = \sqrt{2N}$. The lightest dimension operator is of charge $\pm 1$ and conformal weights $(1/(4N),1/(4N))$. Operators with winding $1$ and momentum $N$ are chiral, and  of conformal weight
$(N,0)$. 
This means that the theory has at least one additional conserved current of spin $N$.
These are therefore rational conformal field theories with an extended W symmetry. These RCFTS have been studied in detail from the point of view of Fusion category symmetry in \cite{Thorngren:2021yso}, which is a more sophisticated version of what we have done here.  Another way to say this, is along the lines of \cite{Chang:2018iay} in section 4.3 is that the non-invertible symmetry we have is the same as the Tambara-Yamagami category for the group ${\mathbb Z}_N$ \cite{tambara1998tensor}. This is because the projector $\Pi$ to the ${\mathbb Z}_N$ invariant projector is given by
\begin{equation}
\Pi= \frac 1 N\sum_k Z^k
\end{equation}
and in the continuum limit one has that $\lim_{L\to \infty} T\simeq 1$ based on how the long wavelength and small energy limit is taken.
This implies that the conformal field theory is equivalent to an orbifold of itself. In this case, the orbifold acts by dividing the radius by $N$. 

Once we have established the conformal field theory, the minimal operator with momentum that is also uncharged under the ${\mathbb Z}_N$ symmetry has conformal weight $(N/4, N/4)$ which is irrelevant when $N>4$. Similar operators of the same conformal weight arise from winding. Indeed, the non-invertible symmetry ${\cal D}$ relates these operators to each other as it acts essentially like T-duality. The non-invertible symmetry therefore predicts additional degeneracies in the boson theory between operators that carry momentum and winding. Notice also that in our analysis we did not make use of the study of defect lines. The defect algebra is how these categorical symmetries are usually presented in the literature, see for example \cite{Aasen:2016dop,Aasen:2020jwb,Zhang:2023wlu} and references therein.

Notice that the relevant operators of the theory all break the  ${\mathbb Z}_N$ symmetry. For $N\leq 4$ we need to assume that the theory is strongly coupled, as the expected deformations of effective field theory from the UV point of view  are either marginal or relevant.
These should induce flows to either a different $c=1$ theory,  or theories with $c<1$ when these deformations are relevant. We seem to see exactly that in the numerical results, which is an a-posteriori obsevation that $c<1$ or that the current algebra is absent.
This also explains why the transition occurs at $N=4$ and why $N>4$ is a locus where we can 
understand the physics directly from the staggered boson variables we have been calling  $q$.

For $N>4$, the other operator that is always there is the $\partial X\bar \partial X$ marginal operator.  
This one has charge $(-1)$ under the non-invertible symmetry ${\cal D}: \bar \partial X\to -\bar \partial X$
so the operator is forbidden if the non-invertible symmetry $\cal D$ is preserved. 
We see that the properties of the critical point at $N>4$ are protected by the ${\mathbb Z}_N$ symmetry and the non-invertible symmetry. 

\subsection{Deformations}

Consider now deforming the clock model theory by modifying the coupling constants in the Hamiltonian, so that 
\begin{equation}
H= H= -\left[\sum_j  g(Q_j+Q^{-1}_j) +(P_{j+1}\otimes P^{-1}_{j}+P_{j}\otimes P^{-1}_{j+1}) \right]
\end{equation}
where we choose $g$ to be real. This choice preserves parity and the ${\mathbb Z}_N$ symmetry. It also  breaks the non-invertible symmetry. Our goal is to understand what happens  when we do this deformation.
We want to argue that this deformation is marginal in the infrared limit.

The idea is that $\partial X\bar\partial X$ is the most relevant operator that is ${\mathbb Z}_N$ invariant and parity invariant. $\partial X$ or $\bar \partial X$ are parity odd and vertex operators $\exp(i kX_L)\exp(i \pm kX_R)$ are $Z_N$ invariant only if $k$ is large enough. As described previously, these start at dimension $(N/4,N/4)$. The upshot is that the most relevant deformation compatible with the symmetries is marginal in the infrared.  

Notice that if we deform the theory by the marginal operator, eventually some of the irrelevant operators that are present in the UV theory at the lattice cutoff can become relevant. This happens because the family of $c=1$ conformal field theories are continuously connected and are related to each other by changing the radius of compactification. When this occurs, one has a transition of infinite order: a BKT transition.
We transition from a gapless phase with continuously varying critical exponents, presumably to a gapped phase with or without symmetry breaking depending on which side of the critical phase we are in.
When we take $g\to 0$, we have $N$ vacua characterized by a state in which $P$ is diagonalized. Because $P$ only appears in the combination $P\otimes P^{-1}$ the eigenvalue of $P$ is not fixed: it will be correlated with the neighbors and this is a global choice. There are $N$ possible values of $P$ associated to the $N$ possible roots of unity that $P$ can evaluate to.
For $g\to\infty$, it is $Q$ that is diagonalized, and we get a trivial vacuum where $Q=1$. We will not consider $g<0$ as it is far away from where our analysis based on the non-ivertible symmetry and perturbation theory will be valid.

This suggests that small deformations of the clock model that break the ${\cal D}$ symmetry but that preserve the ${\mathbb Z}_N$ symmetry
are still critical with $c=1$ at a different radius. That is, the theory is still critical slightly away from the self-dual coupling. This is exactly the phase structure described in  \cite{Elitzur:1979uv} (see also \cite{Fradkin:1980th}).
This type of analysis is very reminiscent of the analysis done in \cite{Jose:1977gm}.

The dimension of the ${\mathbb Z}_{N}$ invariant operators that carry the minimal winding or momentum are given by
\begin{equation}
    (h,\bar h) = \left\{\begin{matrix}  \left( \frac {N R_*^2}{4 R^2},\frac {N R_*^2}{4 R^2} \right) & \hbox{Momentum mode} \\
    \left( \frac {N R^2}{4 R_*^2},\frac {N R^2}{4 R_*^2} \right) &\hbox{Winding mode}
        \end{matrix}\right.
\end{equation}
where $R_*$ is the radius of the circle at the self-dual radius ($\tilde R_*= \sqrt {2N}$). Because we are using ratios, we can equally use $R/R^*=\tilde R/\tilde R^*$ to pass between our conventions of the radius and more standard CFT conventions. 
These become relevant when 
\begin{equation}
    \frac{N R_*^2}{4R^2}=1, \quad \hbox{or}\quad \frac{N R^2}{4R_*^2}=1
\end{equation}
Our goal is now to compute $R/R*$ as a function of the coupling $g$.
We know that $R/R^*=1$ when $g=1$. 

We also notice that the computation of $R*$ we did in equation \eqref{eq:radius} used the $q$ variables directly. We can also use the q-variable directly in the simple approximation that was used to get \eqref{eq:naive}. Our goal is now to combine these two points of view. 

We start with $q$ variables that have the periodicity we have originally described. Namely the one associated to the radius $R_*$. The effective Hamiltonian in the quadratic approximation of the cosines is
\begin{equation}
    H_{eff}= -g \sum_{even}(2-\beta^2 q_{2j}^2) - \sum_{odd}(2-\beta^2 q_{2j+1}^2) 
\end{equation}
Now, we want to rescale the $q$ variables so that  $q_{2j}\to \lambda \tilde q_{2j}$, $q_{2j+1}\to \lambda^{-1} \tilde q_{2j+1}$. This is an automorphism of the Poisson bracket of the $q$ variables.
We do the rescaling so that the coefficient of the quadratic terms in $\tilde q_{2j}^2$ and $\tilde q_{2j+1}^2$ are the same. 
That is
\begin{equation}
    \lambda^2 \beta^2 g = \beta^2 \lambda^{-2}
\end{equation}
That is, $\lambda = g^{-1/4}$. 
So if we assume that the quadratic approximation is reasonable (it is if $\beta$ is very small, which requires $N$ large), then we should be able to read the new periodicity in the $\tilde q$ variables by the rescaling. In this sense, we are allowed to identify
$\lambda \tilde C_{even}= C_{even}$ and $\lambda^{-1} \tilde C_{odd} = C_{odd}$.
The zero mode that is proportional to $C_{even}$ is the one we have been associating to momentum.
That means that when we increase $g$ we are increasing the periodicity of $\tilde C_{even}$ and therefore reducing the radius. Another way to look at it is that we are giving more weight to the energy carried by the even zero mode relative to the odd mode. This increases the energy (dimension) of modes with momentum relative to winding.

With this calculation, the radius is therefore
\begin{equation}
\frac{R^*}{R}= g^{1/4}
\end{equation}
The transition for momentum vertex operators becoming relevant occurs when 
\begin{equation}
\frac{N R_*^2}{4R^2}=1
\end{equation}
so that $g^{(-1/2)}= N/4$.
Similarly, the winding modes become relevant when the radius gets small and
\begin{equation}
\frac{N R^2}{4R_*^2}=1
\end{equation}
so $g^{1/2}= N/4$.
The two BKT transitions are related by $g_1 g_2 =1$, which is a statement about the coupling relations that are required to be able to implement the Kramers-Wanier duality. Remember that in our case of the $q$ variables, this duality is T-duality. 
We expect that a full analysis will have additional perturbative corrections that can modify this quadratic approximation. Understanding such corrections is beyond the scope of the present paper.

Notice also that when $N\to \infty$, the range of $g$ that leads to a critical phase ends up occupying the whole positive real line. This is essentially the theory where we have rescaled $\beta\to 0$ and only the quadratic term in $q$ remains in the effective Hamiltonian. 

Some numerical results on these phase transitions can be found in \cite{Sun:2019ryo}.
They use fidelity susceptibility calculations in DMRG to determine the BKT transitions. 
For $N=5$, we would predict transitions at $g=(5/4)^2$, $g=(4/5)^2$. These are reported at $1.035$ and $0.966$ indicating strong corrections from our naive prediction.
Similarly, for $N=6$ there seem to be large corrections. It would be interesting to understand how that works in detail.

\subsection{A note on quantum circuits}

Consider the problem of trying to simulate the tie evolution of clock models we have described above in terms of a quantum computer with a Troter algorithm. A natural splitting of the problem is to choose
the following splitting for $H=H_1+H_2$ where 
\begin{eqnarray}
    H_1&=& -\left(\sum_j P_j\otimes P_{j+1}^{-1}+P_{j}^{-1}\otimes P_{j+1}\right)\\
    H_2 &=& -\left(\sum_j Q_j+Q_j^{-1}\right)
\end{eqnarray}
and a Trotter step is given by
\begin{equation}
    U= \exp(-i \tau H_1) \exp(-i \tau H_2) =U_1 U_2
\end{equation}
Since all the terms appearing in $H_1$ commute with each other, and all the terms of $H_2$ commute with each other, each of $U_1,U_2$ can easily be implemented on quantum computers with few local qubit gates.
What is the fate of the non-invertible symmetry ${\cal D}$? We clearly see that 
\begin{eqnarray}
    {\cal D} U_1 &=& U_2 {\cal D}\\
    {\cal D} U_2 &=& U_1 {\cal D}
\end{eqnarray}
So we see that ${\cal D}$ does not commute with the unitary operator $U$: it changes the order. 
This would suggest that the non-invertible symmetry is broken. Consider now the expression
\begin{equation}
 {\cal D} U_2 U_1 U_2 = U_1 U_2 {\cal D} U_2
\end{equation}
We see that ${\cal D}U_2$ actually commutes with the unitary $U$. This means that $\tilde {\cal D} = {\cal D} U_2$ should be thought of as the non-invertible symmetry.
The equation that is satisfies is
\begin{equation}
 \tilde {\cal D}^2 = U. {\cal T}\circ \Pi 
\end{equation}
We see that the non-invertible symmetry algebra gets modified non-trivially by the Unitary U.

\section{Conclusion}\label{sec:conc}

In this paper we reviewed the construction of staggered boson variables in a one dimensional lattice.  
We
showed how in the staggered boson variables, the translation symmetry of the lattice by one step is an automorphism of the boson variables that does not automatically translate into a unitary operator acting on Hilbert space. Instead, translation by one step in these variables ends up being given by a non-invertible symmetry which we called ${\cal D}$ (from Duality). The failure of being able to lift the automorphism of the staggered boson variables (one can also say algebra) to a full unitary action on the Hilbert space ends up being entirely determined by the action on  the zero modes of the theory. In the algebra language, these zero modes are the center of the algebra. In contrast, translations by two can always be lifted to unitary operators. 
The translation by one is similar to T-duality. The symmetry algebra that we find is given by
\begin{equation}
    {\cal D}^2 = T\circ \Pi\label{eq:ninv}
\end{equation}
where $\Pi$ is a projector that depends on the allowed values of the zero modes and $T$ is the translation operator (translation by two steps in the staggered boson lattice variables). 

We similarly showed that if we modify the staggered boson algebra to elements $K\simeq \exp(i \beta q)$ that are invariant under gauged translations of the staggered boson variables themselves, then it is possible to write Hamiltonians where the non- invertible symmetry ${\cal D}$ perists and is given by the Kramers-Wannier duality. These Hamiltonians are clock models on the lattice.
In that sense, we showed that the Krammers-Wannier duality can be secretly equivalent to a T-duality transformation.

We analyzed the self-dual point of the clock models by direct diagonalization on small lattices. In these theories, the non-invertible symmetry is unbroken. For clock models ${\mathbb Z}_N$ symmetry with $N>4$, we found that the self-dual point must correspond to a conformal field theory with central charge $c=1$ at radius $\tilde R= \sqrt{2N}$. This result was derived first starting from the staggered boson variables directly. We then argued that this result was exact by using the algebra of the non-invertible symmetry and the relation to T-duality and by verifying the properties on the finite lattice numerical results. 

We also showed that when we modified the coupling away from the self-dual coupling, along a one parameter family of Hamiltonians that preserve the pairty symmetry,  that we should have a critical phase with critical exponents that depend continuously on the coupling and must end in a BKT transition. The boson variables let us estimate 
when the BKT transition should take place. The window of couplings that correspond to  the critical phase grows with $N$. 
This is similar to the behavior of the $XXZ$
model in the critical phase regime. Our result probably has perturbative corrections in the effective radius of the CFT away from the self-dual point that we did not compute. It should be interesting to study these corrections as a computation in the staggerd boson variables as well as numerical  calculations on the lattices.

We also showed that there is a remnant of the non-invertible symmetry if one tries to evolve the system using a Trotter expansion (Floquet dynamics) instead of continuous time evolution. Whether this is useful or not is still to be determined.

Finally, the clock models that we had seem to lead to modular invariant partition functions in the continuum limit. This is not automatic from the point of view of the staggered boson variables. in the staggered bosons variables these conditions are put in by hand. They arise instead of mapping some other variables to the staggered boson variables. It would be interesting to  understand if there is an additional structure that one can add to the staggered boson variables that would make this possibility manifest.

\begin{acknowledgements}
D.B. would like to thank S. Catterall, P. Fendley, Y. Meurice, K. Schoutens and E. Verlinde  for various discussions.
Research supported in part by the Department of Energy under Award No. DE- SC0019139. DB is also supported in part by the Delta ITP consortium, a program of the Netherlands Organisation for Scientific Research (NWO) funded by the Dutch Ministry of Education, Culture and Science (OCW)
\end{acknowledgements}

\begin{appendix}

\section{Finite size effects}\label{app:FS}

Exact diagonalization techniques are limited by the exponential growth of the Hilbert space with increased volume of the lattice $L$. This problem limits our resolution to lattice sizes in this paper $L<10$. Unfortunately the correspondence $H_{\text{CFT}}\sim H_{\text{Lattice}}$ benefits from increasing lattice size and the   equivalence to the CFT occurs only in the limit $L\rightarrow\infty$. In our setup, like any other setup where one is not strictly in the conformal theory,  there will be corrections arising from finite size effects \cite{Cardy:1986ie} which will introduce a discrepancy $\Delta_{\text{CFT}}-\Delta_{\text{Lattice}}=\delta_{\text{F. S.}}$. These finite size effects are evident in the numerics presented. We need some information on their size to estimate the scaling errors and the error bars.

 We use the known finite size behavior \cite{Cardy:1986ie}
\begin{equation}\label{eq:fs}
\begin{aligned}
E_0\approx a_0 L-\frac{a_1 c}{12 L}+\mathcal{O}(L^{-3})\\
E_n-E_0\approx\frac{a_1\Delta_n}{L}\\
E_n-E_m\approx \frac{a_1(\Delta_n-\Delta_m)}{L}
\end{aligned}
\end{equation}
in order to calculate the central charge and the lowest lying primary operator's scaling dimension for the $N=3,4,5$ case. The term $a_0$ above is the energy density of the ground state in the large volume theory.
It appears in every other energy computation. What is usually more relevant is the differences of energies between states. These are all of order $1/L$. The additional $1/L$ correction to the ground state energy is the Casimir of the ground state. This is indicated by the $a_1$ coefficient. It is a physical property of conformal field theories. It depends on the central charge only. The quantity $a_1$ determines the units of energy.
What is more important is that if we want to setup units where the energies of excited states in the cylinder are of order one, then we need to rescale $a_1$.

Because we have used the symbol for approximate signs, we need to do a fit in $1/L$ of the corresponding quantities. We assume that the corrections are of order $1/L^2$ or higher.

A more precise form of these corrections assumes that near the large $L$ limit we can use effective field theory near the conformal point. That is, that the Hamiltonian in the IR scale (the scale $L$) is written as 
\begin{equation}
H_L = H_{CFT} + H_{irr}
\end{equation}
where the $H_{irr}$ are irrelevant corrections which have size of order $1$ at the lattice scale.
We then write the appropriate scaling relations as follows for the Hamiltonian density 
\begin{equation}
{\cal H}_L = {\cal H}_{CFT} + \sum_j L^{2-\Delta_j} a_j {\cal O}_j\label{eq:Hfs}
\end{equation}
with $a_j$ coefficients of order one.
The corrections induced by the operator ${\cal O}_j$ are controlled by its scaling exponent $\Delta_j$. 
At this stage, this scaling is a way to restore the units to CFT units in the infrared (where the energies are of order one).
The point is that corrections to \eqref{eq:fs} are suppressed by inverse powers of L, and that these arise from the scaling dimensions of the irrelevant operators $\Delta_j$. Usually, which precise operators appear as finite size corrections might depend on the symmetries of the system: some irrelevant operators are missing if the lattice theory preserves the symmetries, rather than these being emergent symmetries in which case the corrections would be expected to be present. 
The theory of scaling corrections therefore depends on knowing  the values of $\Delta_i$. Via the operator-state correspondence the dimensions can be correlated with the energies of the excitations on the cylinder. That is, the possible labels  $\Delta_j$ in \eqref{eq:fs} can be mapped to the labels $\Delta_j$ in \eqref{eq:Hfs}. 
If the operator state correspondence holds, then the values of $\Delta_i$ that we compute from energies will inform us directly on the systematic structure of the $1/L$ corrections for other states.

For the purposes of the rest of this appendix, we will use language that assumes the operator 
state correspondence implicitly.

The operator $T\bar T$ of dimension $4$ is in principle always allowed as it is not charged under additional symmetries that the Hamiltonian might posses. If this is the operator that controls the deviations from conformal physics, it will induce corrections that scale as an extra power of $1/L^2$ relative to the correction to the term $a_1 c/L$ which arises from the Cassimir of the field theory at finite volume (essentially if ${\vev H}\sim 1/L $, then $\vev{T}\sim 1/L^2$ and the correction to the density is suppressed by an extra power of $1/L^2$). When integrated we get finite size corrections that go as $1/L^3$ in the un-normalized energies at finite volume. If some irrelevant operators ${\cal  O}_j$ are such that $2<\Delta_j<4$ and are allowed on the right hand side of \eqref{eq:Hfs}, then the finite size effects will be larger (less suppressed at large $L$).

To address the problems numerically, the first step in this process is fitting the ground state energies $E_0$ as a function of lattice size. However this is an unnormalized fit and as such the Casimir term (ie $\frac{c}{12L}$) does not return the correct central charge. In order to normalize the energies of the lattice we must use the properties of the CFT. Since the lattices are relative small, the corrections are still large and can not all be sorted out independently because we do not have enough data for a large 
$L$ fit.

The next problem is to normalize $a_1$ carefully. This is done by knowing a particular difference in energies exactly. The vacuum state $E_0\simeq \ket 0$ has a descendant $L_{-2}\ket 0$ of energy two and spin $2$.  For that state $\Delta =2$, but the finite size corrections can be large as  the finite size corrections scale like $\Delta^2/L^2\sim 4/L^2$ and $\Delta$ is relatively large.
Other primary states $\ket \psi$ have descendants $L_{-1} \ket \psi$ with spin one and energy one higher, so the relative correction is of order $\sim (\Delta_\psi+1)^2/L^2$ which is milder. After all, it is  usually the case that $\Delta_\psi<1$ is small. The best result we get uses the smallest non-trivial primary and we perform the fit to $\Delta E=1$. We can then compare the primary to the ground state and evaluate $\Delta_\psi$. As a consistency, we use $L_{-2}\ket 0$ as a test of how well the system is performing. 

For $N=3$ we use lattice sizes of $L=7,8,9$ and calculate this way the central charge $c=.819$. We then calculate the two lowest lying primary operators to have scaling dimension $0.138$ (with a charge degeneracy) and $0.860$. These correspond to the primaries labeled 2,3 and 4 in Table \ref{tab:Z3} and show close agreement satisfying our check.

For $N=4$ we use lattice sizes of $L=6,7,8$ and calculate the central charge $c=1.038$. We expect the correct central charge to be less than or equal to one, so we can use the deviation from one as a proxy of systematic errors of about $4\%$.
We then calculate the three lowest lying primary operators to have scaling dimensions $0.130$ with a double degeneracy of charge $C=\pm1$ and also $0.260$. These correspond to the primaries labeled $2,3$ and $4$ in Table \ref{tab:Z4} and show close agreement satisfying our check.

For $N=5$ we use lattice sizes of $L=5,6,7$ and calculate the central charge $c=1.08$. We calculate the four lowest primaries to have scaling dimension $.108$ and  $0.361$ both with a double degeneracy of charge. These correspond to the primaries labeled 2,3,4 and 5 in Table \ref{tab:Z5} and show close agreement satisfying our check. 

A point to consider is that the Virasoro symmetry is an emergent property of the large $L$ limit. This means that the Ward identities of the Virasoro algebra also have finite size corrections. These are harder to understand than the description above based on effective field theory. On the other hand, using the Virasoro algebra Ward identities also gives determinations of the representation content of the field theory and we include that information as well.

We believe the various techniques we implement to calculate and validate scaling dimension and conformal dimensions work together to support our conclusions based on exact diagonalization of the theory on small lattices. 

\section{Equivalence of the critical $N=4$ theory and two copies of the critical Ising model}\label{app:twoising}

Our goal in this section is to show that the $N=4$ theory we constructed is equivalent to two copies of 
the critical Ising model in a transverse magnetic field. The idea is that with four states per site, one can map the problem on the site to two qubits.

We  map the states as follows, so that in each step we change at most the value of one bit as follows
\begin{equation}
    \ket 0\to \ket{00}, \ket{1}\to \ket{01}, \ket{2}\to \ket{11}, \ket{3}\to \ket {10}.\label{eq:order}
\end{equation}

We also choose the basis where $P$ is diagonal and equal to 
\begin{equation}
P\sim \hbox{diag} (1, i, -1, -i)
\end{equation}
Since $P$ is diagonal, it must be a combination of the Pauli matrices $1, Z_1, Z_2, Z_1Z_2$ for the two qubits. We normalize $Z$ so that $Z\ket 0= \ket0$ and $Z\ket 1= -\ket 1$. Solving for the linear combination we find that $P$ simplifies to
\begin{equation}
    P \equiv \frac {1-i} 2 Z_1+\frac{1+i}2 Z_2
\end{equation}
When we put this representation in the $P\otimes P^{-1}$ combination, the product becomes
\begin{equation}
P\otimes P^{-1}= \frac 12 (Z_1 \otimes Z_1)-\frac{i}2(Z_1\otimes Z_2)+\frac i2(Z_2\otimes Z_1)+\frac 12 (Z_2\otimes Z_2)
\end{equation}
In the sum $P\otimes P^{-1}+P^{-1}\otimes P$ the imaginary terms cancel and we are left with
\begin{equation}
    P\otimes P^{-1}+c.c= (Z_1 \otimes Z_1)+(Z_2\otimes Z_2)
\end{equation}
Similarly, for $Q$ we use the order defined in \eqref{eq:order} to raise the $\ket n$ label by flipping one bit at a time. 
\begin{eqnarray}
Q&=& X_1\left(\frac{1+Z_1}{2}\right)\left(\frac{1+Z_2}2\right)\nonumber\\
&+&X_2\left(\frac{1-Z_1}{2}\right)\left(\frac{1+Z_2}2\right)\nonumber\\
&+&X_1\left(\frac{1-Z_1}{2}\right)\left(\frac{1-Z_2}2\right)\nonumber\\
&+&X_2\left(\frac{1+Z_1}{2}\right)\left(\frac{1-Z_2}2\right)\nonumber
\end{eqnarray}
In the combination $Q+Q^{-1}$ we get expressions of the form
\begin{equation}
 X_1\left(\frac{1+Z_1}{2}\right)\left(\frac{1+Z_2}2\right)+\left(\frac{1+Z_1}{2}\right)\left(\frac{1+Z_2}2\right)X_1
\end{equation}
and we see that the terms with $Z_1$ appearing here cancel each other as they anticommute with $X_1$. 
A straightforward collection of the terms shows that the result simplifies considerably to
\begin{equation}
    X_1+X_2
\end{equation}
Notice that in this way the Hamiltonian is now written entirely in terms where the $X_1,Z_1$ variables are completely separated from the $X_2,Z_2$ variables. Therefore the Hamiltonian splits $H=H_1+H_2$ with $[H_1,H_2]=0$. Each one of $H_1, H_2$ is a copy of the Ising model in a transverse magnetic field.

In this representation the $Z_4$ symmetry of the $P,Q$ variables is completely obscured.

\end{appendix}
\bibliographystyle{apsrev4-1}[longbibliography]
\bibliography{CFT2.bib}

\end{document}